\documentclass[aps,nofootinbib,prd,twocolumn,preprintnumbers]{revtex4-1}
 \pdfoutput=1
 \usepackage[utf8]{inputenc}

\usepackage[usenames,dvipsnames]{color}  
\usepackage{graphicx}
\usepackage{caption}
\usepackage{subcaption}
\usepackage{amsmath}
\usepackage{amssymb}
\usepackage[colorlinks=true,citecolor=darkred,urlcolor=darkred, pdfborder={0 0 0}]{hyperref}
\usepackage[normalem]{ulem}

\usepackage[T1]{fontenc}
\usepackage{array}
\usepackage{booktabs}
\usepackage{mathrsfs}
\usepackage{multirow}
\usepackage{tabularx}
\usepackage{lipsum}
\usepackage{adjustbox}

\definecolor{darkred}{rgb}{0.6,0,0}

\definecolor{linkcolor}{rgb}{0,0,0.5}

\def\gsim{\raise0.3ex\hbox{$\;>$\kern-0.75em\raise-1.1ex\hbox{$\sim\;$}}}
\def\lsim{\raise0.3ex\hbox{$\;<$\kern-0.75em\raise-1.1ex\hbox{$\sim\;$}}}

\def\beqn#1{\begin{equation}\label{#1}}
\def\eeqn{\end{equation}}

\def\beqa#1{\begin{eqnarray}\label{#1}}
\def\eeqa{\end{eqnarray}}

\def\Z2{$\mathcal{Z_2}$}

\newcommand {\ignore}[1]{}

\def\cevns{CE$\nu$NS~}

\newcommand{\AddrAHEP}{%
  AHEP Group, Institut de F\'{i}sica Corpuscular --
  CSIC/Universitat de Val\`{e}ncia, Parc Cient\'ific de Paterna.\\
 C/ Catedr\'atico Jos\'e Beltr\'an, 2 E-46980 Paterna (Valencia) - Spain}

\begin{document}

\bibliographystyle{unsrt}   

\title{\boldmath \color{BrickRed} COHERENT constraints after the COHERENT-2020 quenching factor measurement}

\author{D.K. Papoulias}\email{dipapou@ific.uv.es}\affiliation{\AddrAHEP}

\begin{abstract}
Recently an improved quenching factor (QF) measurement for low-energy
nuclear recoils in CsI[Na] has been reported by the COHERENT Collaboration. The new energy-dependent QF is characterized by
a reduced systematic uncertainty and leads to a
  better agreement between the experimental COHERENT data and the
  Standard Model (SM) expectation. In this work, we report updated
  constraints on parameters that describe the process of coherent
  elastic neutrino-nucleus scattering within and beyond
  the SM, and we also present how the new QF affects their
  interpretation.
  
\end{abstract}

\maketitle

\section{Introduction}
The first observation of coherent elastic neutrino-nucleus scattering
(CE$\nu$NS) was made at the COHERENT experiment using a CsI[Na]
detector at the Spallation Neutron Source
(SNS)~\cite{Akimov:2017ade,Akimov:2018vzs}, providing a novel powerful
probe for a wide range of low-energy physics searches.
This motivated a large number of theoretical studies to analyze the recorded \cevns signal for performing precision tests of the Standard Model (SM)~\cite{Canas:2018rng} and for investigating possible signatures of new physics beyond the SM~\cite{Lindner:2016wff,Billard:2018jnl,AristizabalSierra:2018eqm,Miranda:2019skf}. The subject became of intense interest during the latest period, and a plethora of extensive studies constantly appear covering a wide spectrum of new physics phenomena such as nonstandard interactions (NSIs)~\cite{Liao:2017uzy,Dent:2017mpr,AristizabalSierra:2017joc,Denton:2018xmq,Dutta:2019eml,Coloma:2017ncl,Gonzalez-Garcia:2018dep},  neutrino  electromagnetic properties~\cite{Kosmas:2015sqa,Kosmas:2017tsq,Miranda:2019wdy,Parada:2019gvy}, sterile neutrinos~\cite{Kosmas:2017zbh,Canas:2017umu,Blanco:2019vyp}, charge-parity violation~\cite{AristizabalSierra:2019ufd}, and new mediators~\cite{Dent:2016wcr,Farzan:2018gtr,Abdullah:2018ykz,Brdar:2018qqj}. Nuclear and atomic effects were also explored in Refs.~\cite{Cadeddu:2017etk,Ciuffoli:2018qem,Huang:2019ene,AristizabalSierra:2019zmy,Papoulias:2019lfi,Arcadi:2019uif,Cadeddu:2019qmv}, which may have direct implications for the neutrino-floor~\cite{Papoulias:2018uzy,Boehm:2018sux,Link:2019pbm} and  dark matter searches~\cite{Ge:2017mcq,Ng:2017aur}. Moreover, from the perspective of experimental physics, several experimental proposals aim to measure \cevns at the SNS~\cite{Akimov:2018ghi} and at reactor facilities~\cite{Hakenmuller:2019ecb,Aguilar-Arevalo:2016qen,Agnolet:2016zir,Billard:2016giu,Strauss:2017cuu,Wong:2010zzc,Belov:2015ufh,Akimov:2016xdc}
(for a review, see Ref.~\cite{Akimov:2019dzb}).

Experiments looking for \cevns and direct dark matter signals are
typically based on accurate measurements of the nuclear response and
are aiming to achieve keV or sub-keV threshold capabilities depending
on the nuclear target. In such measurements, most of the nuclear
recoil energy is dissipated as heat and ionization, while the recorded
energy for the case of scintillator detectors is in reality an
electron equivalent energy whose magnitude depends on the so-called
quenching factor (QF)~\cite{Collar:2014lya}. The QF is an
energy-dependent quantity that is different for a given isotope, and
its calibration involves neutron scattering measurements~\cite{Collar:2013gu}. Regarding the first observation of \cevns at COHERENT with a 14.57~kg CsI[Na]
detector, the first theoretical simulations adopted
an energy-independent QF of $8.78 \pm 1.66 \%$ in the search region
5--30~$\mathrm{keV_{nr}}$~\cite{Scholz:2017ldm}. In this
work we employ the new energy-dependent QF resulted by the COHERENT-2020 campaign~\cite{Konovalov:M7s} from a refined analysis correcting systematic effects of previous measurements, i.e. Chicago-1, Chicago-2, Duke.\footnote{For the Chicago-3 analysis, see Ref.~\cite{Collar:2019ihs}.}

We first show that the new QF measurement leads to a higher
consistency between the SM expectation and the experimental data, a
result that is in agreement with Ref.~\cite{Collar:2019ihs}. We then
revisit various constraints on conventional and exotic parameters
describing the \cevns interaction and update their status. In the
first stage, we explore the sensitivity to the weak mixing angle and
to the average nuclear root-mean-square (rms) radius of CsI assuming purely SM
interactions. Afterward, we reexamine the
sensitivity of COHERENT to phenomenological parameters in the framework of new
physics interaction channels such as vector NSIs, neutrino magnetic
moments, and charge radii as well as in simplified scenarios with novel vector-$Z^\prime$ and scalar mediators. The new  constraints are obtained on the basis of an improved 
$\chi^2$ fit analysis that incorporates the aforementioned quenching
factor effects. We show that the new energy-dependent QF combined with
the reduced uncertainty leads to stronger constraints compared to 
previous studies.

The paper is organized as follows: In Sec.~\ref{sec:simulations} we
provide all necessary ingredients to accurately simulate the observed
\cevns signal, in Sec.~\ref{sec:results} we provide the numerical
results of our sensitivity analysis and update the constraints on the
parameters describing the studied conventional and exotic physics
phenomena. Finally, in Sec.~\ref{sec:conclusions} we summarize the
main outcomes of our work.

\section{Simulation of the COHERENT \cevns rate}
\label{sec:simulations}

During the \cevns interaction, a neutrino with energy $E_\nu$ 
scatters off a nuclear target ($A,Z$) with $Z$
protons and $N=A-Z$ neutrons, which in turn produces a detectable
nuclear recoil $T_A$. Focusing on the COHERENT experiment, after
summing appropriately over the nuclear isotopes $x=\mathrm{Cs, I}$ and
all incident neutrino flavors $\nu_\alpha= (\nu_e, \nu_\mu,
\bar{\nu}_\mu)$, the number of expected \cevns events is given by
\begin{widetext}
\begin{equation}
\begin{aligned}
N_{\mathrm{theor}} = \sum_{\nu_\alpha} \sum_{x = \mathrm{Cs, I}}  N_{\mathrm{targ}}^x  \int_{T_{\mathrm{th}}}^{T_{A}^{\mathrm{max}}} \int_{E_{\nu}^{\mathrm{min}}}^{E_{\nu}^{\mathrm{max}}} f_{\nu_{\alpha}}(E_\nu) \mathcal{ A } ( T_A ) \left(\frac{{d \sigma}_{x}}{dT_A}(E_\nu, T_A) \right)_{\lambda}   dE_\nu dT_A \, ,
\end{aligned}
\label{eq:events}
\end{equation}
\end{widetext}
and depends on the differential cross section $(d
\sigma_x/dT_A)_\lambda$ that is relevant in the framework of a
neutrino interaction channel $\lambda$ within or beyond the
 SM. The number of target nuclei contained in the CsI
detector with mass $m_{\mathrm{det}}=14.57$~kg is determined by Avogradro's number $N_A$ and the stoichiometric ratio $\eta_\chi$ through the relation
$N_{\mathrm{targ}}^x = \frac{m_{\mathrm{det}} \eta_x}{\sum_x A_x
  \eta_x} N_A$. The neutrino-energy flux at the SNS consists
of a prompt and a delayed beam that is adequately described by the
Michel spectrum~\cite{Louis:2009zza}
\begin{equation}
\begin{aligned} 
f_{\nu_\mu}(E_\nu) & = \mathcal{N} \delta\left(E_\nu-\frac{m_{\pi}^{2}-m_{\mu}^{2}}{2 m_{\pi}}\right) \quad &(\text{prompt})\, , \\ 
f_{\bar{\nu}_\mu}(E_\nu) & = \mathcal{N} \frac{64 E^{2}_\nu}{m_{\mu}^{3}}\left(\frac{3}{4}-\frac{E_\nu}{m_{\mu}}\right) \quad &(\text{delayed})\, ,\\ 
f_{\nu_e}(E_\nu) & = \mathcal{N} \frac{192 E^{2}_\nu}{m_{\mu}^{3}}\left(\frac{1}{2}-\frac{E_\nu}{m_{\mu}}\right) \quad &(\text{delayed}) \, ,
\end{aligned}
\label{labor-nu}
\end{equation}
normalized to $\mathcal{N} = r
N_{\mathrm{POT}}/4 \pi L^2$, where $L=19.3$~m is the detector distance from the SNS source, and $r=0.08$ denotes the number of neutrinos per flavor produced for each proton on target (POT), i.e.,
$N_{\mathrm{POT}}=1.76 \times 10^{23}$ for a period of 308.1~days. Assuming SM  interactions, the differential cross section with respect to the nuclear recoil energy is expressed as~\cite{Freedman:1973yd,Papoulias:2015vxa,Bednyakov:2018mjd}
\begin{equation}
\left(\frac{d \sigma}{dT_A}\right)_{\text{SM}} = \frac{G_F^2 m_A}{\pi} (Q_W^V)^2 \left(1 - \frac{m_A T_A}{2 E_\nu^2}\right)  F^2(Q^2) \, ,
\label{eq:xsec-cevns}
\end{equation}
where $m_A$ denotes the nuclear mass and $G_F$ the Fermi coupling constant. The vector $Q_W^V$ weak charge is given by~\cite{Barranco:2005yy} 
\begin{equation}
\begin{aligned}
Q_W^V =  & \left[  2(g_{u}^{L} + g_{u}^{R}) + (g_{d}^{L} + g_{d}^{R}) \right] Z \\ & + \left[ (g_{u}^{L} + g_{u}^{R}) +2(g_{d}^{L} + g_{d}^{R}) \right] N   \, , 
\end{aligned}
\end{equation}
while the $P$-handed couplings of $u$ and $d$ quarks to the
$Z$ boson take the form
\begin{equation}
\begin{aligned}
g_{u}^{L} =& \rho_{\nu N}^{NC} \left( \frac{1}{2}-\frac{2}{3} \hat{\kappa}_{\nu N} \hat{s}^2_Z \right) + \lambda^{u,L} \, ,\\
g_{d}^{L} =& \rho_{\nu N}^{NC} \left( -\frac{1}{2}+\frac{1}{3} \hat{\kappa}_{\nu N} \hat{s}^2_Z \right) + \lambda^{d,L} \, ,\\
g_{u}^{R} =& \rho_{\nu N}^{NC} \left(-\frac{2}{3} \hat{\kappa}_{\nu N} \hat{s}^2_Z \right) + \lambda^{u,R} \, ,\\
g_{d}^{R} =& \rho_{\nu N}^{NC} \left(\frac{1}{3} \hat{\kappa}_{\nu N} \hat{s}^2_Z \right) + \lambda^{d,R} \, .
\end{aligned}
\end{equation}
Here, $\hat{s}_Z^2 \equiv \sin^2 \theta_W= 0.2382$ is the weak mixing angle and $\rho_{\nu N}^{NC} = 1.0082$, $\hat{\kappa}_{\nu N} = 0.9972$, $\lambda^{u,L} = -0.0031$, $\lambda^{d,L} = -0.0025$, and
$\lambda^{d,R} =2\lambda^{u,R} = 3.7 \times 10^{-5}$ are the radiative corrections~\cite{Tanabashi:2018oca}. Because of their tiny contributions to the \cevns rate, axial-vector interactions, incoherent interactions, as well as contributions due to the sodium dopant of the CsI[Na] detector are neglected.

The main source of theoretical uncertainty in the SM \cevns process arises from the  nuclear form factor that takes into account the finite nuclear size and depends on the variation of the momentum transfer $Q^{2}=2 m_A T_A$~\cite{Papoulias:2019lfi}. Following the COHERENT Collaboration, in this work we adopt the Klein-Nystrand (KN) form factor parametrized as~\cite{Klein:1999qj}
\begin{equation}
F_{\text{KN}} = 3 \frac{j_1(Q R_A)}{Q R_A} \left[ 1 + (Q a_k )^2 \right]^{-1} \, ,
\end{equation}
where $a_k = 0.7$ fm is the range of the Yukawa potential (over a Woods-Saxon distribution) in the hard sphere approximation with radius $R_A= 1.23 \times A^{1/3}$.  We note that regarding the old QF, slight differences from the corresponding results of Ref.~\cite{Kosmas:2017tsq} throughout the paper are due to the adoption of the KN form factor, the different neutrino-energy distribution considered, the different value of the weak mixing angle, as well as the binned $\chi^2$ analysis performed here (see below).

For a scintillation-based experiment, the  measured quantity is the number of photoelectrons (PEs) denoted here as $n_\mathrm{PE}$. To account for this mechanism, the \cevns  differential rate in events vs nuclear recoil  energy gets converted to an equivalent differential rate in events vs  electron recoil energy through the application of the QF function $Q_f(T_A)$, and that in turn gets converted to a PE spectrum via the light yield $\mathcal{L}_Y= 13.348\, \mathrm{PE/keV_{ee}}$ measured for electron recoils as 
\begin{equation}
n_\mathrm{PE} =  Q_f(T_A) \mathcal{L}_Y T_A\, .
\label{eq:PE_vs_Tnr}
\end{equation}
In Eq.(\ref{eq:events}), the acceptance efficiency of the CsI detector is taken into account, which in terms of the photoelectron content  of the signal reads\footnote{Note that the efficiency function is instrumental and does not depend on the QF.}~\cite{Akimov:2018vzs}
\begin{equation}
\mathcal{A}(n_\mathrm{PE}) = \frac{k_1}{1 + e^{- k_2 \left(n_\mathrm{PE} - x_0 \right)}} \Theta(n_\mathrm{PE}) \, ,
\label{eq:eff-SNS}
\end{equation}
with $k_1= 0.6655$, $k_2= 0.4942$, $x_0= 10.8507$, and the modified Heaviside function 
\begin{equation}
\Theta(n_\mathrm{PE}) = \left\{ \begin{array}{ll}{0} & {n_\mathrm{PE} < 5} \\ {0.5} & {5 \leq n_\mathrm{PE} < 6} \\ {1} & {n_\mathrm{PE} \geq 6 \, .} \end{array} \right.
\end{equation}
\begin{figure}[t]
\includegraphics[width= 0.5 \textwidth]{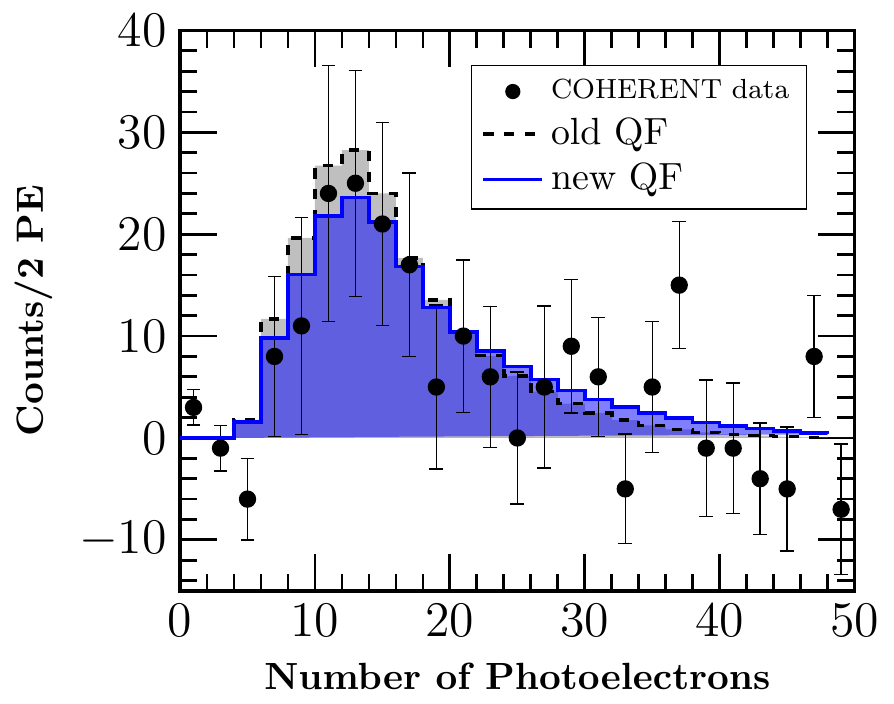}
\caption{Comparison of the expected number of events at the COHERENT CsI detector for the old vs the new QF measurement.}
\label{fig:events}
\end{figure}

Up to now, previous analyses adopted the 
  energy-independent QF of $8.78 \pm 1.66 \%$ recommended by the
  COHERENT Collaboration in Ref.~\cite{Akimov:2017ade} which carried a large uncertainty of 25\%.\footnote{In reality, the QF uncertainty is $18.9\%$ leading to an overall uncertainty in \cevns rate of 25\%~\cite{communication}. We however adopt the official values reported in Ref.~\cite{Akimov:2017ade}.} In the present work we consider the new energy-dependent QF
which came out of the refined COHERENT-2020 measurement
with a reduced uncertainty by about a factor of 4 at  3.6\% (for more
details, see Ref.~\cite{Konovalov:M7s}). In agreement with
Ref.~\cite{Konovalov:M7s}, within the SM the new calculation gives a
theoretical value of  158 events as compared to the $174$ events
corresponding to the old QF. At this point, it is rather important to emphasize that a better agreement is now reached with the 134 events observed in
Ref.~\cite{Akimov:2017ade}. The corresponding results are compared in Fig.~\ref{fig:events} as a function of the PE bins.

\section{Numerical Results}
\label{sec:results}

\begin{figure*}[t]
\includegraphics[width= \textwidth]{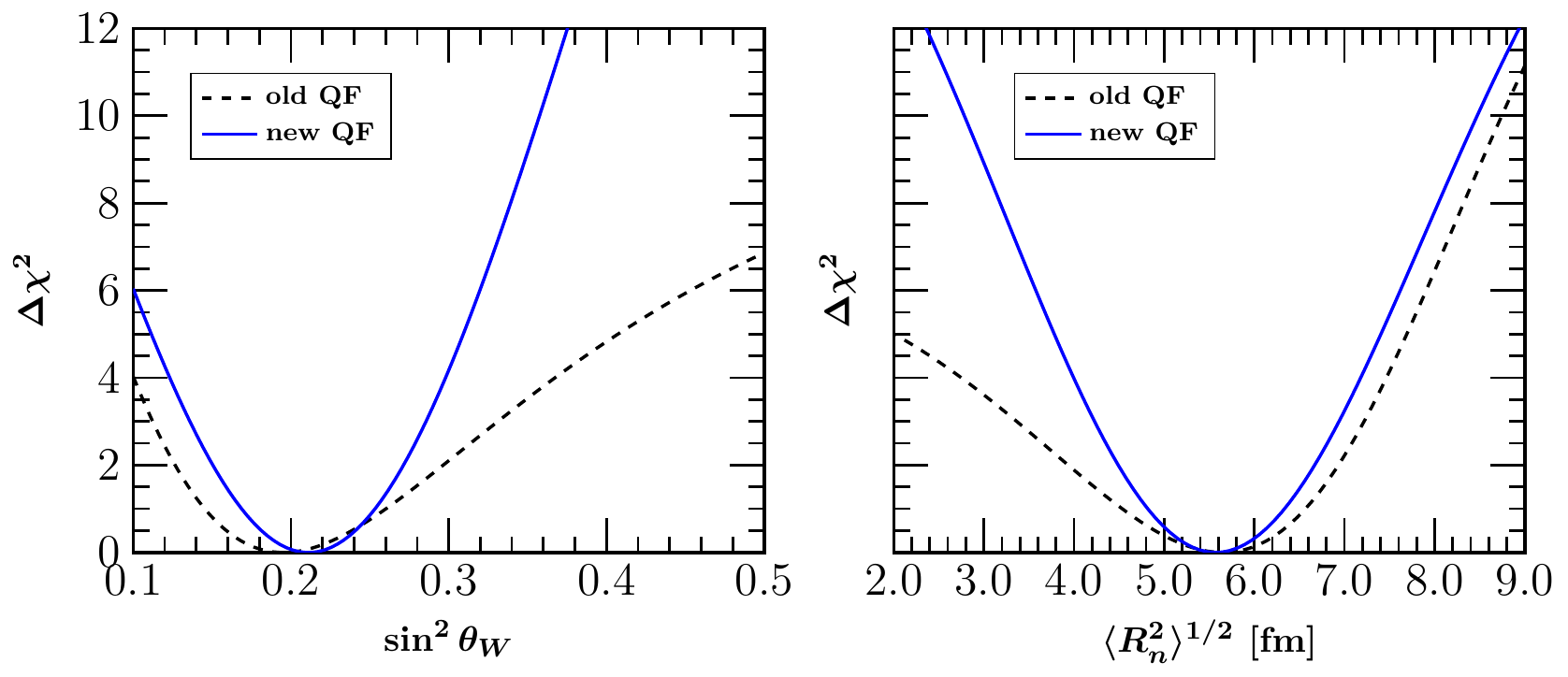}
\caption{$\chi^2$ profiles for the cases of the weak mixing  angle (left) and the average  nuclear rms radius of CsI (right) extracted from the analysis of the COHERENT data for the old vs the new QF measurement.}
\label{fig:SM}
\end{figure*}

In the present study, we perform a sensitivity analysis of the parameter set $\mathcal{S}$ in question (see below) that follows from a $\chi^2 (\mathcal{S})$ fit that is relevant for the CsI detector at the COHERENT experiment and reads~\cite{Akimov:2017ade} 
\begin{widetext}
\begin{equation}
\begin{aligned}
\chi^2 (\mathcal{S}) =  \underset{\mathtt{a}_1, \mathtt{a}_2}{\mathrm{min}} \Bigg [ \sum_{i=4}^{15} \left(\frac{N^i_{\mathrm{meas}} - N^i_{\mathrm{theor}}(\mathcal{S}) [1+\mathtt{a}_1] - B^i_{0n} [1+\mathtt{a}_2] }{\sqrt{N^i_{\mathrm{meas}} + B^i_{0n} + 2 B^i_{ss}}}\right)^2 
  + \left(\frac{\mathtt{a}_1}{\sigma_{\mathtt{a}_1}} \right)^2 + \left(\frac{\mathtt{a}_2}{\sigma_{\mathtt{a}_2}} \right)^2 \Bigg ] \, .
\end{aligned}
\label{eq:chi}
\end{equation}
\end{widetext}
Here, $N^i_{\mathrm{meas}}$ ($N^i_{\mathrm{theor}}$) represents the $i$th bin of the observed signal (theoretical CE$\nu$NS events), and $B^i_{0n}$ ($B_{ss}^i$) denotes the beam-on prompt neutron (steady-state) background events taken from the COHERENT data release \cite{Akimov:2018vzs}, while the analysis is restricted to the 12 energy bins corresponding to $6 \leq n_\mathrm{PE} \leq 30$.  In Eq.(\ref{eq:chi}), $\mathtt{a}_1$
and $\mathtt{a}_2$ are the corresponding systematic parameters with
fractional uncertainties  $\sigma_{\mathtt{a}_1} = 12.8\%$ (5\% from
signal acceptance determination, 5\% from form factor choice, 10\%
from neutrino flux and  3.6\% from the new QF) and
$\sigma_{\mathtt{a}_2} = 25\%$. Note that compared to $\sigma_{\mathtt{a}_1}=28\%$
given in Ref.~\cite{Akimov:2017ade} and adopted by all similar studies
up to now, the fractional uncertainty considered here is reduced by
about a factor of 2.  This
is also in agreement with estimations of previous studies addressing possible future experimental
setups~\cite{Cadeddu:2017etk,Papoulias:2019lfi,Miranda:2019wdy} and
will have a direct impact on the updated constraints presented below.

\subsection{SM precision tests and nuclear physics}

Assuming purely SM interactions, we first extract the new sensitivity to the weak mixing angle that arises from the new QF measurement. To this end, we evaluate the $\chi^2(\sin^2 \theta_W)$ function and perform a sensitivity fit by varying around the central value $\sin^2 \theta_W = 0.2382$. The resultant sensitivity profiles are depicted in the left panel of Fig.~\ref{fig:SM}. A comparison with the corresponding result assuming the old energy-independent QF is also shown. Indeed, this new calculation leads to reasonably improved results. From the fit, we find the following constraints  at 90\% C.L.
\begin{equation}
\begin{aligned}
\sin^2 \theta_W =& 0.197^{+0.124}_{-0.080} \quad  \text{(old QF)} \, ,\\
\sin^2 \theta_W =& 0.209^{+0.072}_{-0.069} \quad \text{(new QF)} \, .
\end{aligned}
\end{equation}
Evaluating the $1\sigma$ bands $\delta s_W^2$ according to the definition given in Ref.~\cite{Kosmas:2015sqa}, we find the values $\delta s_W^2=(0.057, 0.042)$ for the (old, new) QF case, which yield the corresponding percentage uncertainties $\delta s_W^2/ \sin^2\theta_W$ of (29\%, 20\%).

We then make an effort to explore the sensitivity to the nuclear rms radius that follows from the recent COHERENT measurement. To this purpose, we employ the refined QF resulting from Ref.~\cite{Konovalov:M7s}, while in this case we consider the Helm form factor~\cite{Helm:1956zz}
\begin{equation}
F_{\text{Helm}}(Q^2) = 3 \frac{j_1(Q R_0)}{q R_0}\, e^{-(Q s)^2/2} \, ,
\label{eq:helm}
\end{equation} 
where $j_1(x)$ is the spherical Bessel function of the first
kind. Here, $\left\langle R^2_n \right\rangle^{1/2} = \sqrt{\frac { 3
  } { 5 } R _ { 0 } ^ { 2 } + 3 s ^ { 2 }}$ is the nuclear rms radius,
$R_0 = 1.23 \, A^{1/3}$~fm is the diffraction radius, and $s=0.9$~fm
quantifies the surface thickness (for more details, see
Refs.~\cite{AristizabalSierra:2019zmy,Papoulias:2019lfi}). The
resultant sensitivity profile is presented in the right panel of Fig.~\ref{fig:SM}, showing that the constraints are 
now stronger than  previously reported~\cite{Cadeddu:2017etk,Ciuffoli:2018qem,Papoulias:2019lfi}. In
particular at 90\% C.L. we find the best fits~\footnote{Note that, in this case the form factor uncertainty is neglected in Eq.(\ref{eq:chi}).}
\begin{equation}
\begin{aligned}
\left\langle R^2_n \right\rangle^{1/2} =& 5.6^{+1.5}_{-2.1} \, \text{fm} \quad  \text{(old QF)} \, ,\\
\left\langle R^2_n \right\rangle^{1/2} =& 5.6^{+1.3}_{-1.25} \, \text{fm} \quad \text{(new QF)} \, .
\end{aligned}
\end{equation}
In a similar manner, within $1 \sigma$ error we find the bands $\delta \left\langle R^2_n \right\rangle^{1/2}=(1.01, 0.76)$ and the corresponding percentage uncertainties (18\%, 14\%) for the (old, new) QF measurement.
We finally stress that the latter results remain essentially the same when considering the Klein-Nystrand form factor.

\subsection{Nonstandard interactions}
\label{sec:NSI}

\begin{figure*}[t]
\includegraphics[width=\textwidth]{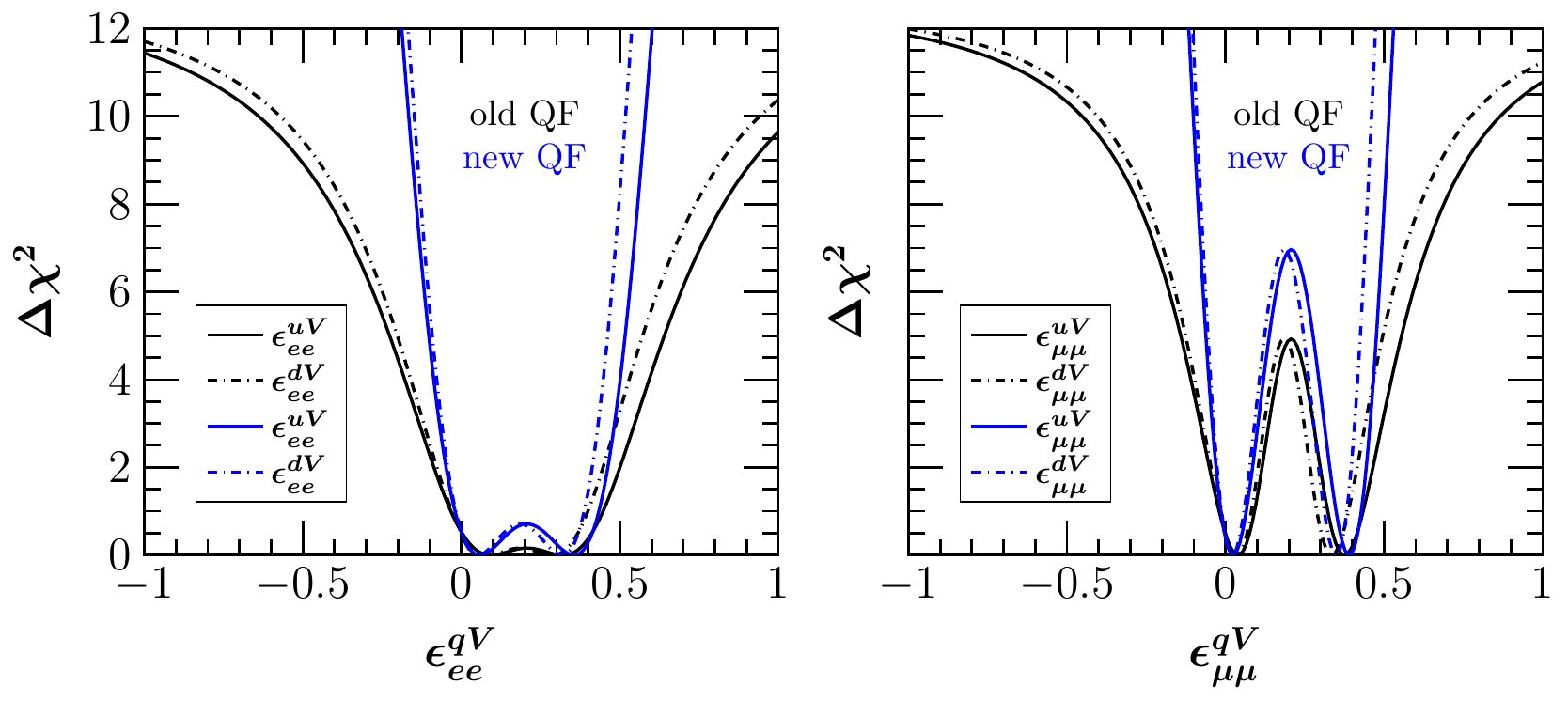}
\caption{$\chi^2$  profiles for the NU NSIs from the analysis of the COHERENT data. A comparison of the obtained sensitivity  using the old vs the new QF is also shown.}
\label{fig:NSI}
\end{figure*}

\begin{figure*}[t]
\includegraphics[width=\textwidth]{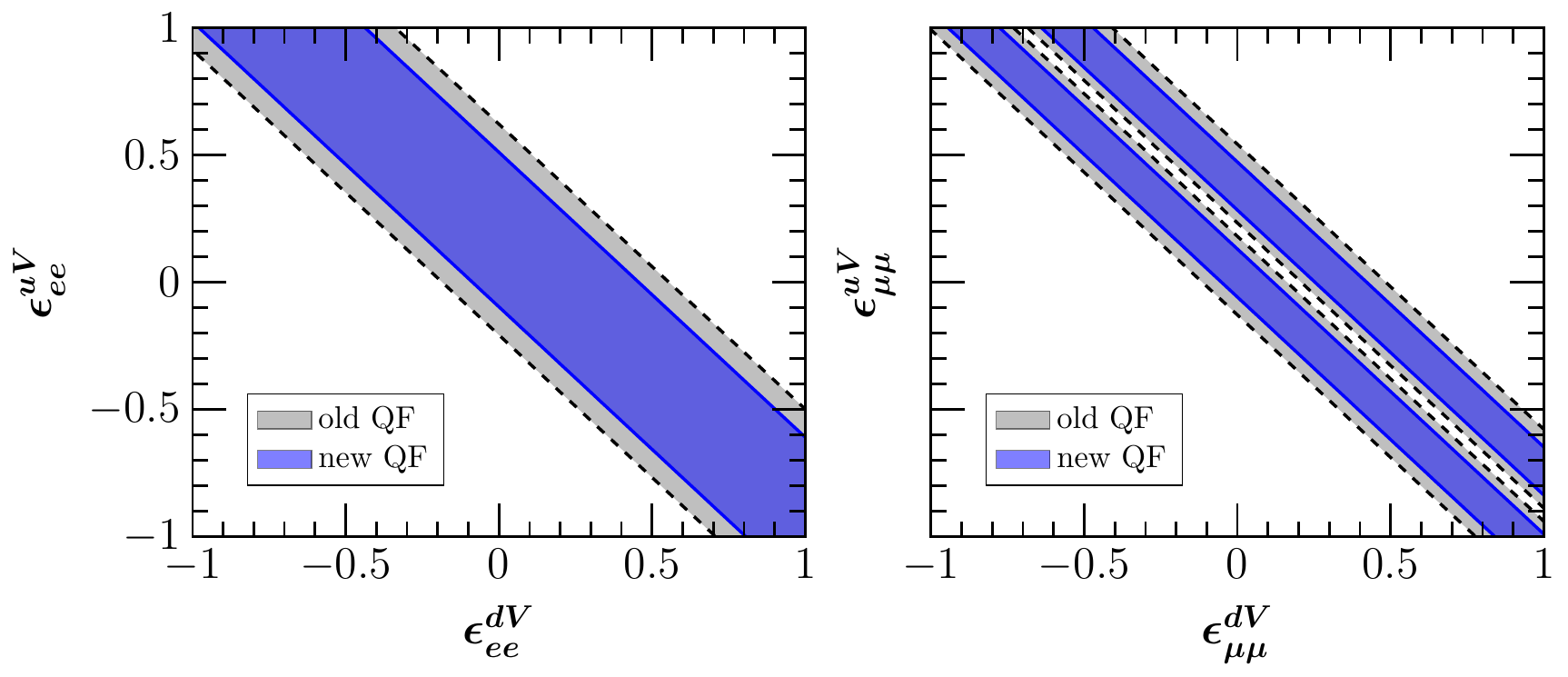}
\caption{Allowed regions in the NU NSIs parameter space obtained from the analysis of the COHERENT data for the old vs the new QF measurement.}
\label{fig:NSI_contour}
\end{figure*}

Nonstandard interactions have been a popular subject of extensive
research during the last 15 years, with interesting applications in
neutrino oscillations and low-energy neutrino physics (for a review, see Refs.~\cite{Miranda:2015dra,Farzan:2017xzy}). For a neutrino with flavor
$\alpha=\{e,\mu,\tau\}$ and a quark $q=\{u,d\}$, the vector-type NSI
contributions that arise due to nonuniversal (NU) flavor-preserving and flavor-changing interactions are described in the NSI weak
charge~\cite{Barranco:2005ps,Scholberg:2005qs}
\begin{equation}
\begin{aligned}
Q_{\mathrm{NSI}}^V = & (2 \epsilon_{\alpha \alpha}^{uV} + \epsilon_{\alpha \alpha}^{dV} + g^V_p) Z + (\epsilon_{\alpha \alpha}^{uV} + 2 \epsilon_{\alpha \alpha}^{dV} + g^V_n) N \\
& + \sum_{\alpha, \beta} \left[ (2 \epsilon_{\alpha \beta}^{uV} + \epsilon_{\alpha \beta}^{dV}) Z + (\epsilon_{\alpha \beta}^{uV} + 2 \epsilon_{\alpha \beta}^{dV} ) N \right] \, .
\end{aligned}
\end{equation}
In the context of NSI, the expected \cevns rate is modified according
to the substitution $Q_W^V \rightarrow
Q_{\mathrm{NSI}}^V$ in the SM differential cross section of
Eq.(\ref{eq:xsec-cevns}).

Assuming a single nonvanishing NSI parameter at a time,
Fig.~\ref{fig:NSI} illustrates the obtained sensitivity for
the NU $\epsilon_{ee}^{qV}$ ($\epsilon_{\mu \mu}^{qV}$) couplings in
the left (right) panel, while a useful comparison is also given for
the case of the old QF.  The
  impact of the new QF measurement on NSI constraints becomes evident.  The left and
right panels of Fig.~\ref{fig:NSI_contour} show the allowed regions at
90\% C.L. in the $(\epsilon_{ee}^{dV},\epsilon_{ee}^{uV})$ and the
$(\epsilon_{\mu \mu}^{dV},\epsilon_{\mu \mu}^{uV})$ parameter space,
respectively.  We see that the bounds are now more
  restrictive than the corresponding results using the old QF.

\begin{figure}[t]
\includegraphics[width=\linewidth]{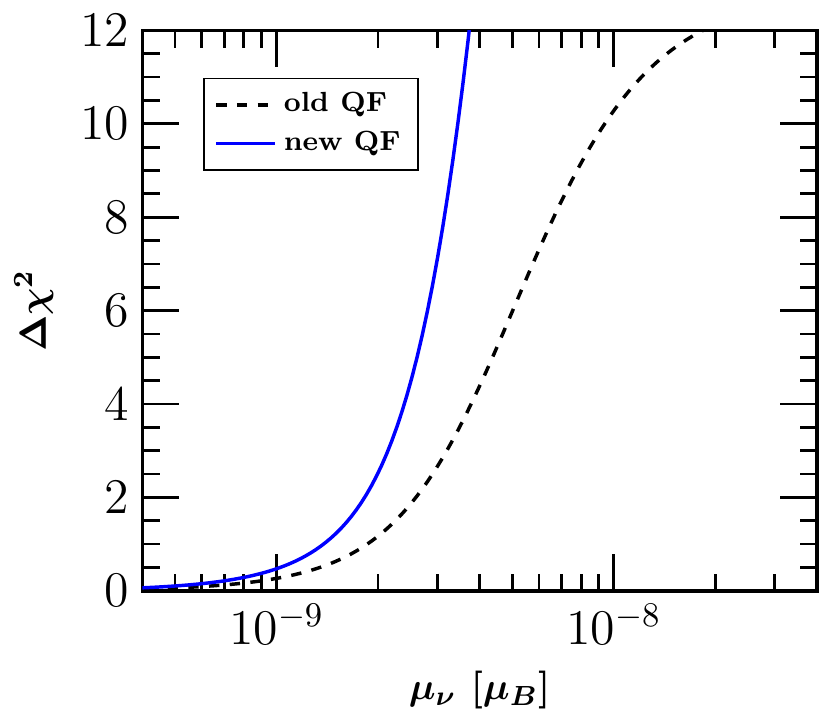}
\caption{$\chi^2$  profiles of the effective
    neutrino magnetic moment extracted by the COHERENT data for the old vs the new QF measurement.}
\label{fig:magnetic}
\end{figure}

\begin{figure*}[t]
\includegraphics[width=\textwidth]{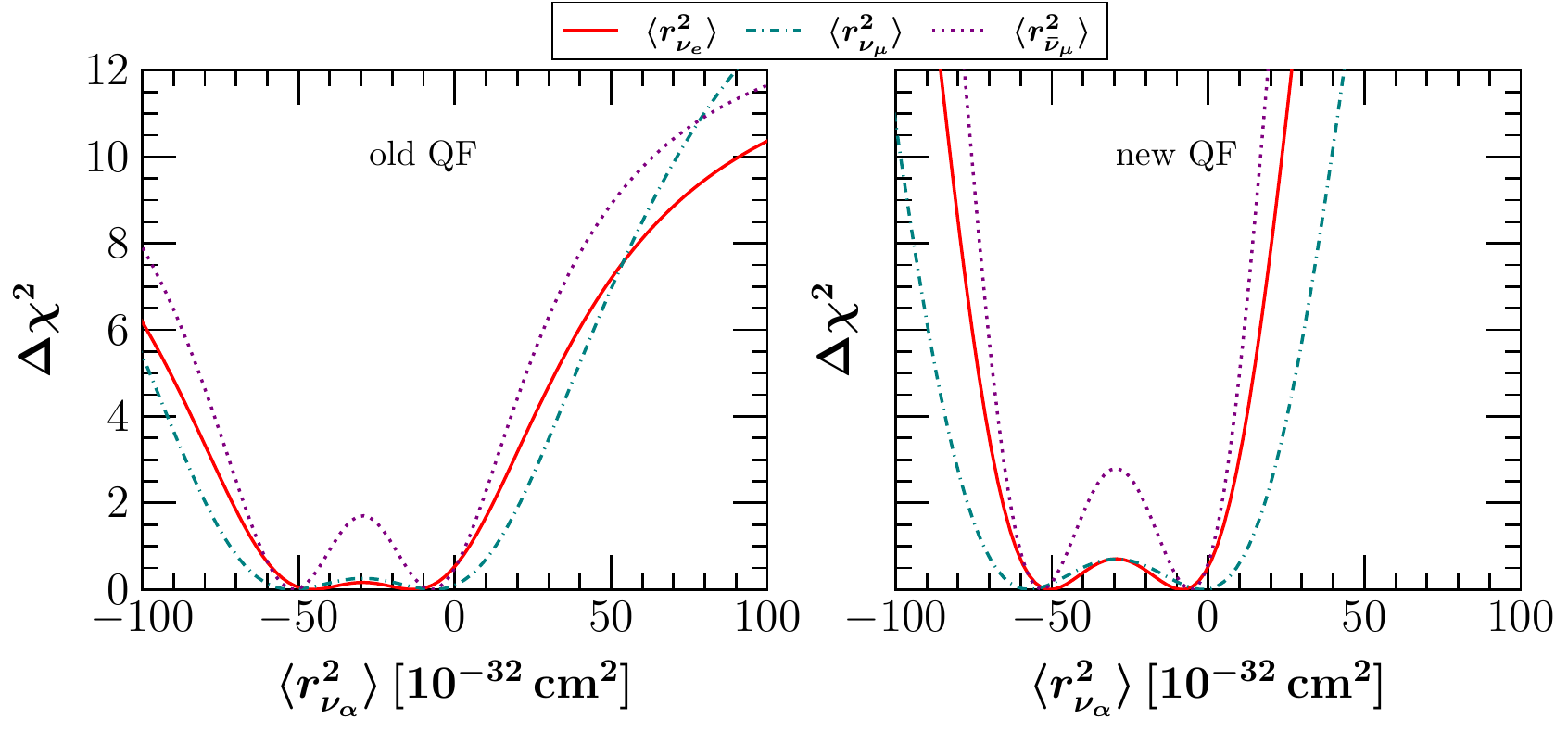}
\caption{$\chi^2$  profiles of the relevant neutrino charge radii at the COHERENT experiment for the old (left) and the new (right) QF measurement.}
\label{fig:rv}
\end{figure*}

\begin{figure*}[t]
\includegraphics[width=\textwidth]{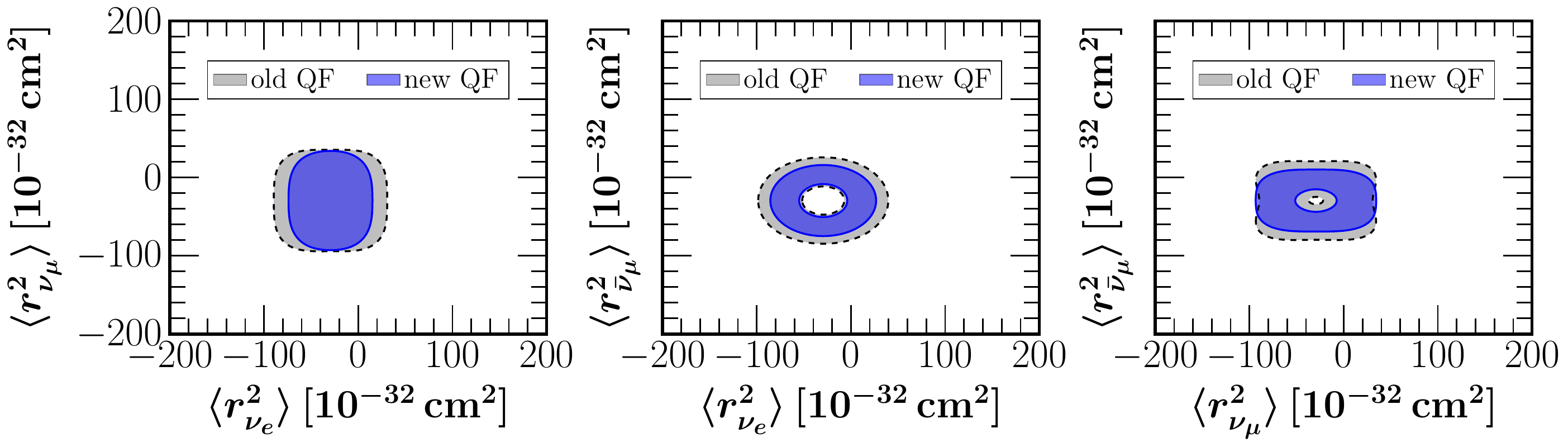}
\caption{Contours in the neutrino charge radius parameter space from
  the analysis of the COHERENT data. The results are shown for various combinations in the $(\langle r_{\nu_\alpha}^2 \rangle, \langle r_{\nu_\beta}^2 \rangle)$ plane and compared for the old vs the new QF measurement.}
\label{fig:rv_contour}
\end{figure*}

\subsection{Electromagnetic neutrino interactions}
In this subsection, we are interested in exploring the possibility of probing nontrivial neutrino electromagnetic (EM) properties~\cite{Schechter:1981hw} and to revisit existing constraints  from \cevns~\cite{Kosmas:2017tsq}. The two main phenomenological parameters that arise in the framework of EM neutrino interactions are the neutrino magnetic moment and the neutrino charge radius. For completeness, we mention that in the simplest Majorana neutrino case, the neutrino magnetic moment $\mu_\nu$ is in reality expressed in terms of the neutrino transition magnetic moments $\Lambda_i$ of the neutrino magnetic moment matrix~\cite{Grimus:2000tq,Tortola:2004vh}, while constraints have been recently extracted from neutrino-electron scattering~\cite{Canas:2015yoa} and \cevns~\cite{Miranda:2019wdy}.  Here, for simplicity, we consider the effective neutrino magnetic moment in the helicity-violating EM cross section~\cite{Vogel:1989iv}
\begin{equation}
\left( \frac{d \sigma}{dT_A} \right)_{\mathrm{EM}}=\frac{\pi a^2_{\text{EM}} \mu_{\nu}^{2}\,Z^{2}}{m_{e}^{2}}\left(\frac{1-T_A/E_{\nu}}{T_A}\right) F^{2}(Q^{2})\,.
\label{NMM-cross section}
\end{equation}
In Fig.~\ref{fig:magnetic}, we present the updated constraint on $\mu_\nu$ from our analysis with the new QF, which is also compared to the corresponding one that comes  from the old QF. The obtained upper limits at 90\% C.L. read
\begin{equation}
\begin{aligned}
\mu_\nu  < & \, 3.1 \times 10^{-9} \, \mu_B \quad  \text{(old QF)} \, ,\\
\mu_\nu  < & \, 2.6 \times 10^{-9} \, \mu_B \quad \text{(new QF)} \, .
\end{aligned}
\end{equation}
From the same plot, it can be deduced that this difference is more pronounced at 99\% C.L.

For a flavor neutrino $\nu_\alpha$, the associated neutrino charge
radius $\langle r_{\nu_\alpha}^2 \rangle$ is another interesting
phenomenological quantity which arises from the helicity-conserving
charge form factor of the EM neutrino current~\cite{Giunti:2014ixa}.
The latter leads to a shift of the weak mixing angle as
follows~\cite{hirsch:2002uv}:
\begin{equation}
\sin^2 \theta_W \rightarrow \sin^2 \overline{\theta_W} + \frac{\sqrt{2} \pi a_{\mathrm{EM}}}{3 G_F} \langle r_{\nu_\alpha}^2 \rangle \, .
\label{eq:charge-radius}
\end{equation}
We stress  that there is not a sign flip regarding antineutrino charge radii; e.g., it holds $\langle r_{\bar{\nu}_\alpha}^2 \rangle=\langle r_{\nu_\alpha}^2 \rangle$ as defined in Ref.~\cite{Kosmas:2017tsq}\footnote{Reference~\cite{Cadeddu:2018dux} used a negative sign which is now corrected in Ref.~\cite{Cadeddu:2019eta}.}. In this work, we follow the definition given in Ref.~\cite{Kosmas:2017tsq} however, the shift considered here is smaller by a factor of 2.
Neglecting transition charge radii and varying one parameter at a time, Fig.~\ref{fig:rv} shows the $\chi^2$ profiles of the neutrino charge radii $\langle r_{\nu_\alpha}^2 \rangle$ associated with the respective SNS neutrino flux, where the left (right) panels correspond to the old (new) QF measurement. The obtained constraints differ slightly due to the old vs new QF data. The only noticeable difference is that by employing the new QF in the case of the prompt $\bar{\nu}_\mu$ beam, the resultant constraint on $\langle r_{\bar{\nu}_\mu}^2 \rangle$ is separated into two distinct regions at 90\% C.L. It is now worthwhile to explore the simultaneous constraints that can be obtained. Figure~\ref{fig:rv_contour} presents the allowed regions at 90\% C.L. in the $(\langle r_{\nu_\alpha}^2 \rangle, \langle r_{\nu_\beta}^2 \rangle)$ parameter space.  As expected the allowed parameter space in all cases is more restricted using the new QF data.

\subsection{Simplified scenarios with light mediators}
In addition to the NSIs discussed previously in Sec.~\ref{sec:NSI}, we are now interested in simplified scenarios where the NSI is generated due to the presence of novel mediators.

\begin{figure*}[ht!]
\includegraphics[width=\textwidth]{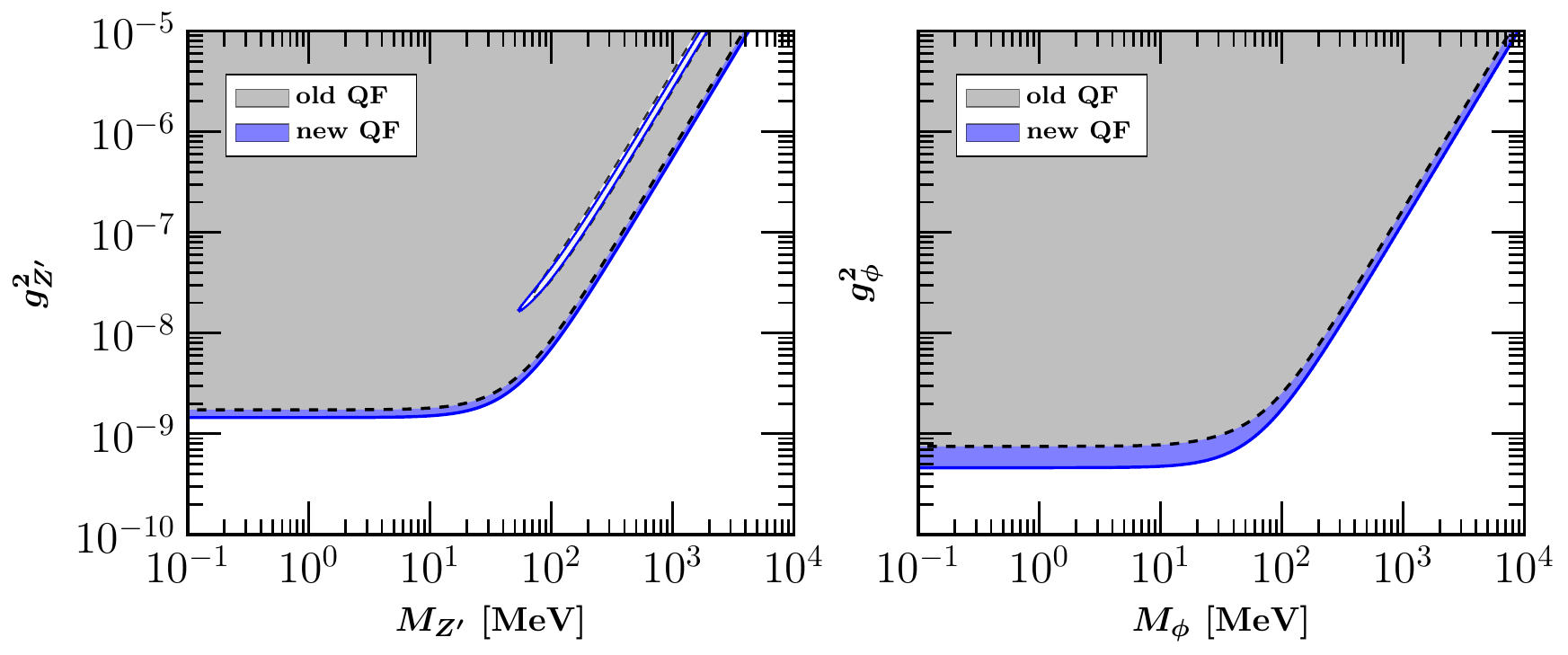}
\caption{Exclusion curves in the $(g^2_{Z^\prime}, M_{Z^\prime})$  parameter space (left) and in the $(g^2_\phi, M_\phi)$  parameter space (right) from the analysis of the COHERENT data. The results are shown for the old and the new QF measurement.}
\label{fig:mediators}
\end{figure*}

In the first step, we explore the case where the \cevns rate is
enhanced from contributions due to a vector-$Z^\prime$ mediator with
mass $M_{Z^\prime}$. The relevant cross section takes the
form~\cite{Bertuzzo:2017tuf}
\begin{equation}
\left( \frac{d \sigma}{dT_A}\right)_{\mathrm{SM} + Z^\prime} = \mathcal{G}_{Z^\prime}^2 (T_A,g_{Z^\prime},M_{Z^\prime}) \left(\frac{d \sigma}{d T_A} \right)_{\mathrm{SM}} \, ,
\end{equation}
with the $Z^\prime$ factor defined as
\begin{equation}
\mathcal{G}_{Z^\prime} = 
1 + \frac{1}{\sqrt{2}G_F}\frac{Q_{Z^\prime}}{Q_W^V} \frac{g_{Z^\prime}^{\nu V}}{2 m_A T_A + M_{Z^\prime}^2} \, .
\label{eq:G_z-prime}
\end{equation}
In the above expression, in order to reduce the number of model parameters, we consider the generalized coupling  $g_{Z^\prime}^2 = g^{\nu V}_{Z^\prime} Q_{Z^\prime}/3 A$  that is expressed in terms of the vector $\nu_{\alpha}$-$Z^\prime$ coupling times the respective vector charge $Q_{Z^\prime}^V$, under the assumption of universal quark-$Z^\prime$ couplings (for more details, see Ref.~\cite{Kosmas:2017tsq}).

Concentrating our attention on the case of a new scalar boson $\phi$ mediating the \cevns process, the cross section takes the form~\cite{Cerdeno:2016sfi}
\begin{equation}
\left(\frac{d \sigma}{d T_A} \right)_{\mathrm{scalar}} = \frac{G_F^2 m_A^2}{ 4 \pi} \frac{\mathcal{G}_\phi^2 M_\phi^4 T_A }{E_\nu^2 \left( 2 m_A T_A + M_\phi^2 \right)^2} F^2(Q^2) \, ,
\end{equation}
with the corresponding scalar factor being 
\begin{equation}
\mathcal{G}_\phi = \frac{g^{\nu S}_\phi Q_\phi}{G_F M_\phi^2} \, .
\end{equation}
In the same spirit of the discussion made above, for the sake of simplification our calculations involve the generalized scalar coupling $g_\phi^2 =  g^{\nu S}_\phi Q_\phi/ \left(14 A + 1.1 Z \right) $.\footnote{This result derives from the nuclear charge related to the scalar boson exchange; see Ref.~\cite{Kosmas:2017tsq}.}

The exclusion regions in the parameter space $(M_{Z^\prime},g_{{Z^\prime}}^2)$ and $(M_\phi,g_\phi^2)$ for the vector and scalar scenarios, respectively, are obtained from a two parameter analysis of the COHERENT data. For both old and new QF data, the results are presented at 90\% C.L. in the left (right) panel of  Fig.~\ref{fig:mediators} for vector (scalar) mediators. As in all previous cases, from this plot we conclude that the new QF data lead to generally more stringent bounds.

\section{Conclusions}
\label{sec:conclusions}
\begin{table*}[ht]
\resizebox{0.5\textwidth}{!}{ 
\begin{tabular}{@{}lcccc@{}}
\toprule
\textbf{parameter}                    & \textbf{old QF}       &   &     & \textbf{new QF}               \\ \midrule
$\sin^2 \theta_W$                     & 0.116 -- 0.321      &     &     & 0.140 -- 0.282                \\
$\langle R_n^2\rangle^{1/2} $          & 3.5 -- 7.1       &     &        & 4.3 -- 6.7                    \\
$\epsilon_{ee}^{uV}$                  & -0.12 -- 0.53       &    &      & -0.06 -- 0.48                 \\
$\epsilon_{ee}^{dV}$                  & -0.11 -- 0.48      &     &      & -0.06 -- 0.43                 \\
$\epsilon_{\mu \mu}^{uV}$             & -0.07 -- 0.13 \& 0.28 -- 0.49 & & & -0.04 -- 0.1 \& 0.32 -- 0.45 \\
$\epsilon_{\mu\mu}^{dV}$              & -0.06 -- 0.12 \& 0.25 -- 0.43 & &  & -0.03 -- 0.09 \& 0.28 -- 0.40 \\
$\mu_\nu$                             & 30              &    &         & 21                           \\
$\langle r_{\nu_e}^2\rangle$          & -76 -- 17         &    &        & -68 -- 9                     \\
$\langle r_{\nu_\mu}^2 \rangle$       & -84 -- 25        &      &       & -80 -- 21                     \\
$\langle r_{\bar{\nu}_\mu}^2 \rangle$ & -71 -- 12      &   &            & -65 -- -32 \& -27 -- 6        \\ \bottomrule
\end{tabular}
}
\caption{Summary of constraints at 90\% C.L. in the present work. The results are extracted assuming the old and the new QF data. The nuclear rms radius is in units of fm, the effective neutrino magnetic moment in  $10^{-10} \, \mu_B$ and the neutrino charge radius in $10^{-32} \, \mathrm{cm^2}$.}
\label{tab:summary}
\end{table*}

Focusing on the COHERENT experiment, we  reexamined the results implied from CE$\nu$NS in light of a new  QF measurement~\cite{Konovalov:M7s}. By using the new QF data, we came out with improved constraints regarding all the cases analyzed in this work. A full summary  is given in Table~\ref{tab:summary}.  At first, we presented updated constraints focusing on important SM parameters namely, the weak mixing angle and the average nuclear rms radius of CsI, and we explicitly demonstrated the level of improvement. We  then concentrated on interesting phenomenological parameters beyond the SM and presented updated constraints for nonuniversal NSIs as well as for electromagnetic neutrino properties including the effective  neutrino magnetic
moment and the neutrino charge radius. Finally, we  revisited the sensitivity of COHERENT in the framework of simplified scenarios involving massive vector and scalar mediators. We concluded that a substantial improvement on SM parameters is reached, while the improvement of beyond the SM physics constraints is also evident.  


\begin{acknowledgments}
The author acknowledges K. Scholberg and J. Collar for useful correspondence. The author is indebted to M. T\'ortola and O. Miranda for their critical comments and their help during the preparation of the manuscript.
This work is supported by the Spanish Grants No. SEV-2014-0398 and No. FPA2017-85216-P (AEI/FEDER, UE), Grant No. PROMETEO/2018/165 (Generalitat
Valenciana) and the Spanish Red Consolider MultiDark Grant No. FPA2017-90566-REDC.

\end{acknowledgments}



\providecommand{\href}[2]{#2}\begingroup\raggedright\endgroup

\end{document}